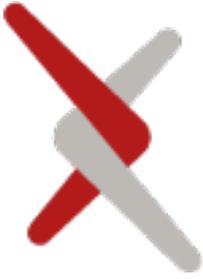

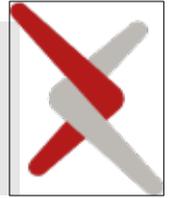

# ArteryX: Advancing Brain Artery Feature Extraction with Vessel-Fused Networks and a Robust Validation Framework

Abrar Faiyaz[a,∗], Nhat Hoang[b], Giovanni Schifitto[a,c,d], Md Nasir Uddin[a,d,e,∗]

[a]*Department of Neurology, University of Rochester, Rochester, NY 14642, USA*
[b]*Department of Physics, University of Rochester, Rochester, NY 14627, USA*
[c]*Department of Imaging Sciences, University of Rochester, Rochester, NY 14642, USA*
[d]*Department of Electrical & Computer Engineering, University of Rochester, Rochester, NY 14627, USA*
[e]*Department of Biomedical Engineering, University of Rochester, Rochester, NY 14627, USA*

## ABSTRACT

Cerebrovascular pathology significantly contributes to cognitive decline and neurological disorders, underscoring the need for advanced tools to assess vascular integrity. Three-dimensional Time-of-Flight Magnetic Resonance Angiography (3D TOF MRA) is widely used to visualize cerebral vasculature, however, clinical evaluations generally focus on major arterial abnormalities, overlooking quantitative metrics critical for understanding subtle vascular changes. Existing methods for extracting structural, geometrical and morphological arterial features from MRA - whether manual or automated - face challenges including user-dependent variability, steep learning curves, and lack of standardized quantitative validations. We propose a novel semi-supervised artery evaluation framework, named ArteryX, a MATLAB-based toolbox that quantifies vascular features with high accuracy and efficiency, achieving processing times ∼10–15 minutes per subject at 0.5 mm resolution with minimal user intervention. ArteryX employs a vessel-fused network based landmarking approach to reliably track and manage tracings, effectively addressing the issue of dangling/disconnected vessels. Validation on human subjects with cerebral small vessel disease demonstrated its improved sensitivity to subtle vascular changes and better performance than an existing semi-automated method. Importantly, the ArteryX toolbox enables quantitative feature validation by integrating an *in-vivo* like artery simulation framework utilizing vessel-fused graph nodes and predefined ground-truth features for specific artery types. Thus, the ArteryX framework holds promise for benchmarking feature extraction toolboxes and for seamless integration into clinical workflows, enabling early detection of cerebrovascular pathology and standardized comparisons across patient cohorts to advance understanding of vascular contributions to brain health.

## 1. Introduction

The cerebral vasculature plays a pivotal role in maintaining brain function, and its dysfunction is increasingly recognized as a key contributor to cognitive decline and dementia, even without overt brain infarctions (Wardlaw et al., 2019). Populations with vascular risk factors–such as hypertension, diabetes, dyslipidemia, smoking, HIV, chronic kidney disease, obesity, metabolic syndrome, obstructive sleep apnea, systemic autoimmune diseases like lupus, or exposure to cranial irradiation or neurotoxic chemotherapy–exhibit accelerated cerebral small vessel disease (CSVD) (Toyoda, 2015; Dearborn et al., 2015; Schammel et al., 2022; Wiseman et al., 2016; Remes et al., 2020). These conditions drive endothelial dysfunction, chronic inflammation and vascular rarefaction, disrupting neural connectivity and impairing cognitive performance (van Dinther et al., 2022).

Understanding and quantifying the intracranial vascular network is critical for early detection of CSVD, elucidating the impact of aging and disease on cerebral arteries, and guiding timely therapeutic interventions to mitigate cognitive decline (Elahi et al., 2023). Furthermore, current clinical evaluations, such as those using Three-dimensional Time-of-Flight Magnetic Resonance Angiography (3D TOF MRA), primarily focus on large vessel stenoses or aneurysms, often rely on subjective visual assessments rather than precise quantitative measurements (De Koning et al., 2003; Hsu et al., 2017). These asssessments are reported to have inter- and intra-annotator variability and is referred to as weak labeling for validation(Tan et al., 2022; Vos et al., 2025). Yet every automated and semi-automated techniques we have found in the literature has been validated with weak labeling, exposing the need for a robust validation framework to accurately characterize the discrete artery features.(Dumais et al., 2022; Chen et al., 2018; Nader et al., 2023; Vos et al., 2025)

### 1.1. Related Works

After segmentation of the TOF-MRA data, two classes of problems are observed in the quantification of intracranial artery profiles from 3D TOF-MRA.

Firstly, the classification problem: Intracranial artery anatomical configurations vary widely. For example, (Lippert and Pabst, 1985) categorize the anterior and posterior Circle of Willis (CoW) into 10 classes each, while (Lazorthes et al., 1979) propose 22 variants. More recent studies have introduced simplified and complex versions of these classifications (Chen et al., 2019, 2018; Dumais et al., 2022; Nader et al., 2023), leading to inconsistent comparisons across studies (Vos et al., 2025). Most of these approaches focus solely on the proximal CoW – except iCafe(Chen et al., 2018) – and tend to ignore distal artery quantification due to the complexity in those regions in terms of annotator variability, and processing efficiency.

Secondly, feature extraction and its grounded validation for discrete arteries: Commonly extracted features from TOF-MRA – such as artery radii, tortuosity, length and related morphological features – are often based on centerline extractions of the overall intracranial artery system, which limits the evaluation of arteries discretely (Deshpande et al., 2021; Shahzad et al., 2015). Additionally, state-of-the-art tools like iCafe require significant expert time (∼1.5 hours per subject), while deep learning approaches demand large training datasets (e.g., over 100 subjects), making them impractical for clinical evaluations.(Chen et al., 2018; Shahzad et al., 2015; Nader et al., 2023; Dumais et al., 2022). These approaches also lack quantification of structured topology and complexity feature.

In short, automated techniques are typically trained and evaluated on weakly labeled data, overlook distal arteries, and lack standardization in classification and feature extraction. While semi-automated approaches such as iCafe offer certain level of standardization, they fall short in capturing vascular topology and complexity, and are inefficient in processing subjects.

### 1.2. Contributions

This study introduces a novel data structure to address the limitations of existing methods for intracranial artery quantification, offering the following contributions in a toolbox published online named 'ArteryX':

1. *Efficient Semi-Automated and Quantitative Vascular Analysis*: We proposed a robust, non-invasive method to quantitatively characterize the intracranial vascular network using 3D TOF MRA, enabling precise measurement of morphological, complexity and topological features for discretely classified distal and proximal arteries of CoW.

2. *ArteryX Validation Framework*: We generated a 3D vascular graph with connected tracings proposed as vessel-fused data structure. This data structure allows for defining artery specific diameters and orientations to simulate ground-truth features for method validation. To the best of our knowledge, our framework is the first that resembles the complexity of in-vivo data observed in 3D TOF-MRA.

3. *Standardized and Generalizable Intracranial Artery Classification*: We addressed the lack of standardization in artery classification, building on top of the artery classification framework in iCafe(Chen et al., 2018), where we included topological characterization and generalization of distal arteries using of sub-network reporting.

4. *Enhanced Clinical Applicability*: By significantly reducing processing time compared to iCafe and eliminating the need for extensive manual intervention, our approach supports practical integration into clinical workflows. Furthermore, its demonstrated application in a CSVD group comparison with iCafe indicates enhanced sensitivity compared to the state-of-the-art.

―――――――――
*Corresponding Authors:
Abrar Faiyaz (abrar_faiyaz@urmc.rochester.edu) and Md Nasir Uddin (nasir_uddin@urmc.rochester.edu)

---

**Algorithm 1** ArteryX Workflow

---

1: **function** ARTERYX
2:     ArteryBinary ← EM(*Parameters*)                        ▷ Estimate binary artery segmentation
3:     Skeleton3D ← CenterlineExtraction(ArteryBinary)          ▷ Extract 3D arterial skeleton
4:     VesselFusedNetwork ← VesselFusedNetworkCreation(Skeleton3D)    ▷ Create vessel-fused graph
5:     PreparedGraph ← PrepareForLabeling(VesselFusedNetwork, HessianGuide)    ▷ Prepare graph with Hessian guide
6:     Landmarks ← WaitForUserLandmarks(PreparedGraph, max = 16)    ▷ User identifies up to 16 main landmarks
7:     DynamicGraphTable ← CreateDynamicGraphTable(PreparedGraph, Landmarks)    ▷ Account for missing arteries
8:     ArteryFeatures ← SegmentedArteryFeatureExtraction(DynamicGraphTable)    ▷ Extract artery features
9:     **return** ArteryFeatures, DynamicGraphTable
10: **end function**

---

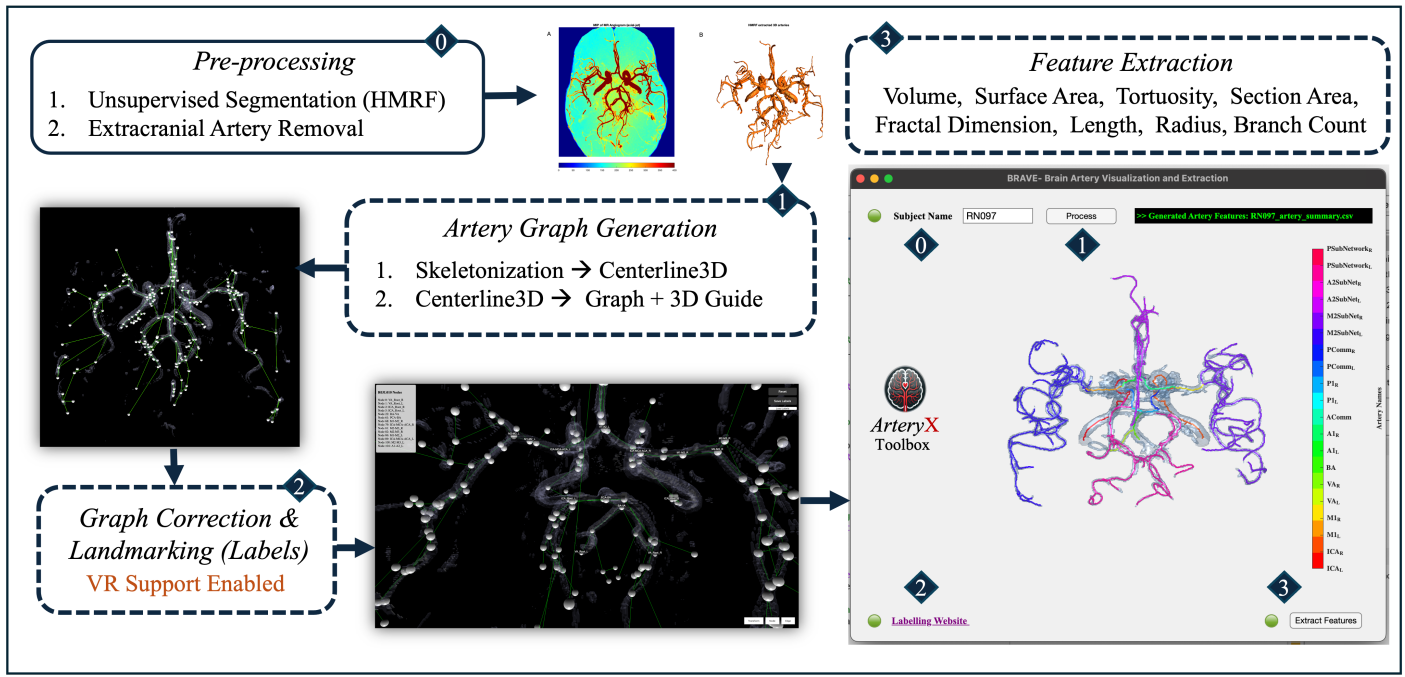

Fig. 1: ArteryX Evaluation Pipeline for arterial centerline extraction and analysis.

## 2. Methods

### 2.1. ArteryX Estimation Framework

Fig. 1 illustrates the proposed ArteryX pipeline for intracranial artery segmentation, landmarking, and vascular feature extraction. The detailed steps of the ArteryX pipeline are outlined in Algorithm 1. The unsupervised segmentation stage is the first cornerstone of the ArteryX estimation pipeline, enabling automated extraction of arterial centerlines and their corresponding radial measurements. This process is followed by isotropic rasterization. During preprocessing, raw arterial data is first segmented using Hidden Markov Random Field - Expectation Maximization (HMRF-EM) to seperate arterial structures from surrounding tissues. The HMRF-EM framework leverages probabilistic modeling to optimize an energy function that balances image intensity, spatial coherence, and contextual dependencies, effectively handling noise and low-contrast regions in medical imaging data Wang (2012). Key variables in HMRF include pixel/voxel intensities, neighborhood configurations, Gaussian mixture model parameters (means, variances, and weights), and iterative optimization factors such as conver-gence thresholds. These variables enable robust segmentation of arterial boundaries, providing a clean binary image for extracting the vessel-fused network and subsequently relevant arterial features.

#### 2.1.1. The Expectation-Maximization for Unsupervised Artery Segmentation

The Expectation-Maximization (EM) Algorithm (Zhang et al., 2001) is a robust statistical method used for parameter estimation in probabilistic models with latent variables. In the context of medical image segmentation, the EM is employed to refine the segmentation of a corrupted image by iteratively estimating class memberships and updating model parameters. To enhance spatial coherence, the EM algorithm integrates Iterated Conditional Modes (ICM) for maximum a posteriori (MAP) estimation, applied to a segmented image with a background mask. The algorithm-2 optimizes the segmentation by balancing data likelihood and spatial priors, controlled by a tunable penalty parameter.

The EM algorithm alternates between two steps: the Expectation (E) step, which computes the expected class member-ships for each pixel, and the Maximization (M) step, which updates the model parameters (class means and standard deviations) based on these memberships. The implementation incorporates ICM to refine the segmentation by enforcing spatial consistency, guided by a Markov Random Field (MRF) prior. The algorithm terminates when the relative change in the log-posterior probability falls below a threshold or when the posterior decreases, indicating convergence or a suboptimal update.

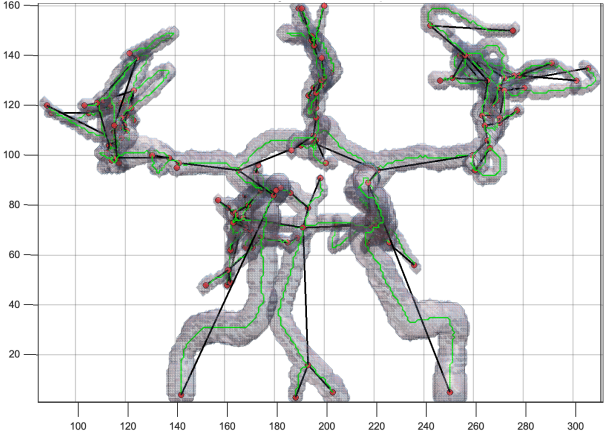

Fig. 2: Vessel-Fused Artery Graph, illustrating graph edges and nodes, and arterial pathways (traces).

The inputs to the algorithm include:

- **img**: The original 3D MRA image.

- **seg**: The initial segmented image, where each pixel is assigned a class label. (Prior is calculated from an initial constant threshold)

- **mask**: A binary background mask, where 1 indicates foreground pixels.

- $k$: The number of classes. (Fixed as 2, foreground and background)

- $\mu$: Initial class means, $\mu_i$ for class $i \in \{1, \ldots, k\}$.

- $\sigma$: Initial class standard deviations, $\sigma_i$ for class $i$.

- $\beta$: A tunable penalty parameter for the MRF prior. (Fixed as 1)

- $\epsilon_{\text{EM}}$: Stopping criterion for the EM algorithm.

- $N_{\text{ICM}}$: Number of ICM iterations. (Configured based on dataset SNR-level, for our overall dataset we fixed it at 10)

- $N_{\text{EM max}}$: Maximum number of EM iterations. (Configured based on dataset SNR, for our overall dataset it was fixed at 4)

The outputs are the updated segmented image **seg**, class memberships **M**, and updated parameters $\mu$ and $\sigma$.

The EM algorithm seeks to maximize the log-posterior probability of the segmentation given the observed image, which combines a likelihood term and a prior term. For a pixel at position $(r, c)$, the observed intensity is $y_{r,c} = \textbf{img}(r, c)$, and the latent class label is $z_{r,c} \in \{1, \ldots, k\}$. The likelihood assumes a Gaussian distribution for each class:

$$p(y_{r,c}|z_{r,c} = i, \mu_i, \sigma_i) =$$
$$\mathcal{G}(y_{r,c}; \mu_i, \sigma_i) = \frac{1}{\sqrt{2\pi\sigma_i^2}} \exp\left(-\frac{(y_{r,c} - \mu_i)^2}{2\sigma_i^2}\right), \quad (1)$$

where $\mathcal{G}(y; \mu, \sigma)$ denotes the Gaussian probability density function.

The prior term is modeled using an MRF, which encourages spatial smoothness by penalizing dissimilar labels in neighboring pixels. The prior probability for a pixel $(r, c)$ belonging to class $i$ is

$$p(z_{r,c} = i|\textbf{seg}, \beta) \propto \exp\left(-\beta \cdot U_{r,c}(i, \textbf{seg})\right), \quad (2)$$

where $U_{r,c}(i, \textbf{seg})$ is the prior penalty, typically the number of neighboring pixels with different labels, weighted by $\beta$. The log-posterior probability can be computed as-

$$\log p(\textbf{seg}|\textbf{img}, \textbf{mask}, \mu, \sigma, \beta) =$$
$$\sum_{r,c:\textbf{mask}(r,c)=1} \Big[ \log \mathcal{G}(\textbf{img}(r, c); \mu_{z_{r,c}}, \sigma_{z_{r,c}})$$
$$- \beta U_{r,c}(z_{r,c}, \textbf{seg}) \Big] \quad (3)$$

The EM algorithm iteratively updates the segmentation and parameters to maximize this posterior.

*E-Step: Compute Class Memberships*

In the E-step, the algorithm computes the posterior probability (class membership) of each pixel $(r, c)$ belonging to class $i$, given the current parameters and segmentation. For foreground pixels ($\textbf{mask}(r, c) = 1$), the unnormalized membership is-

$$m_{r,c,i} = \mathcal{G}(\textbf{img}(r, c); \mu_i, \sigma_i) \cdot \exp\left(-\beta U_{r,c}(i, \textbf{seg})\right). \quad (4)$$

The memberships are normalized to ensure they sum to 1 over all classes:

$$\textbf{M}(r, c, i) = \frac{m_{r,c,i}}{\sum_{j=1}^{k} m_{r,c,j}}. \quad (5)$$

Before computing memberships, the algorithm updates the segmentation using ICM, which iteratively assigns each pixel the label that maximizes the local posterior probability, considering both the likelihood and the MRF prior. ICM runs for $N_{\text{ICM}}$ iterations, improving the MAP estimate of **seg**.

*M-Step: Update Parameters*

In the M-step, the class means and standard deviations are updated using the memberships from the E-step. For each class $i$, the updated mean is the weighted average of pixel intensities:

$$\mu_i = \frac{\sum_{r,c} \textbf{M}(r, c, i) \cdot \textbf{img}(r, c)}{\sum_{r,c} \textbf{M}(r, c, i)}, \quad (6)$$

where the summation is over all pixels $(r, c)$.

The updated variance is computed as:

$$\sigma_i^2 = \frac{\sum_{r,c} \mathbf{M}(r,c,i) \cdot (\mathbf{img}(r,c) - \mu_i)^2}{\sum_{r,c} \mathbf{M}(r,c,i)}, \tag{7}$$

and the standard deviation is $\sigma_i = \sqrt{\sigma_i^2}$.

*Convergence Check*

The algorithm evaluates convergence by computing the log-posterior probability before and after the ICM update. The relative change is:

$$\text{rel\_change} = \frac{|\log p_{\text{after}} - \log p_{\text{before}}|}{|\log p_{\text{before}}|}. \tag{8}$$

The algorithm terminates if $\text{rel\_change} \leq \epsilon_{\text{EM}}$, if the posterior decreases ($\log p_{\text{after}} < \log p_{\text{before}}$), or if the maximum number of iterations $N_{\text{EM max}}$ is reached.

The algorithm's convergence is sensitive to the choice of $\epsilon_{\text{EM}}$ and $\beta$. A small $\epsilon_{\text{EM}}$ ensures precise convergence but may increase computational cost, while a large $\beta$ enforces stronger smoothing, potentially oversmoothing fine details. The maximum iteration limit $N_{\text{EM max}}$ prevents excessive computation in cases of slow convergence.

The EM algorithm has been widely used in medical image segmentation, often combined with HMRF-EM priors to enforce spatial constraints (Zhang et al., 2001), in our implementation a binary thresholding artery image initiates the EM and ICM iterations. The incorporation of ICM for MAP estimation is inspired by Julian(Besag, 1986), who introduced ICM as an efficient alternative to global optimization methods like simulated annealing. For a specific group of dataset that is obtained from the same scanner and protocol, that demonstrates similar SNR, we recommend to keep all the tunable parameters including the beta and number of iterations fixed. This addresses the inherent bias that originates from the using different contrast algorithms or parameters for segmentation.

### 2.1.2. *Isotropic Rasterization & Geodesic Skeletonization*

Subsequently, the segmented data is standardized into an isotropic grid through resampling and rasterization, ensuring uniform voxel spacing for consistent analysis. The centerline and radius computation establishes a robust foundation for constructing a graph representation of the arterial network, critical for topological and quantitative vascular analyses.

1. **Input Preparation:** The algorithm processes the 3D binary image where a unit structure specifying voxel dimensions along each axis ($x$, $y$, $z$). The voxels representing the arterial structure are resampled to isotropic voxel space to account for anisotropy, ensuring that distance calculations are physically accurate and corrected for the imaging system's spatial units.

2. **Distance Transform Computation:** A distance transform is applied to the inverted binary image to compute the Euclidean distance from each voxel within the arterial structure to the nearest background voxel (i.e., close to the vessel wall). In the implementation, the `bwdist` function in MATLAB computes the Euclidean distance

---

**Algorithm 2** EM Algorithm for Artery Segmentation

---

1: **function** EM(**seg**, **img**, **mask**, $k$, $\mu$, $\sigma$, $\beta$, $\epsilon_{\text{EM}}$, $N_{\text{ICM}}$, $N_{\text{EM max}}$)
2:　　Initialize iteration counter $i_{\text{em}} \leftarrow 0$
3:　　**while** true **do**
4:　　　　Compute $\log p_{\text{before}} \leftarrow \log p(\mathbf{seg}|\mathbf{img}, \mathbf{mask}, \mu, \sigma, \beta)$
5:　　　　Update $\mathbf{seg} \leftarrow \text{ICM}(\mathbf{seg}, \mathbf{img}, \mathbf{mask}, k, \sigma, \beta, N_{\text{ICM}})$
6:　　　　Compute $\log p_{\text{after}} \leftarrow \log p(\mathbf{seg}|\mathbf{img}, \mathbf{mask}, \mu, \sigma, \beta)$
7:　　　　Compute $\text{rel\_change} \leftarrow \frac{|\log p_{\text{after}} - \log p_{\text{before}}|}{|\log p_{\text{before}}|}$
8:　　　　**if** $\text{rel\_change} \leq \epsilon_{\text{EM}}$ **or** $\log p_{\text{after}} < \log p_{\text{before}}$ **then**
9:　　　　　　**break**
10:　　　**end if**
11:　　　Initialize membership array $\mathbf{M} \leftarrow \text{zeros}(R, C, k)$
12:　　　**for** $r = 1$ to $R$ **do**
13:　　　　　**for** $c = 1$ to $C$ **do**
14:　　　　　　　**if** $\text{mask}(r, c) = 1$ **then**
15:　　　　　　　　　**for** $i = 1$ to $k$ **do**
16:　　　　　　　　　　　Compute $m_{r,c,i} \leftarrow \mathcal{G}(\mathbf{img}(r,c); \mu_i, \sigma_i) \cdot \exp(-\beta U_{r,c}(i, \mathbf{seg}))$
17:　　　　　　　　　**end for**
18:　　　　　　　　　Normalize $\mathbf{M}(r, c, i) \leftarrow \frac{m_{r,c,i}}{\sum_{j=1}^{k} m_{r,c,j}}$
19:　　　　　　　**end if**
20:　　　　　**end for**
21:　　　**end for**
22:　　　**for** $i = 1$ to $k$ **do**
23:　　　　　Compute $\mu_i \leftarrow \frac{\sum_{r,c} \mathbf{M}(r,c,i) \cdot \mathbf{img}(r,c)}{\sum_{r,c} \mathbf{M}(r,c,i)}$
24:　　　　　Compute $\sigma_i^2 \leftarrow \frac{\sum_{r,c} \mathbf{M}(r,c,i) \cdot (\mathbf{img}(r,c) - \mu_i)^2}{\sum_{r,c} \mathbf{M}(r,c,i)}$
25:　　　　　Set $\sigma_i \leftarrow \sqrt{\sigma_i^2}$
26:　　　**end for**
27:　　　Increment $i_{\text{em}} \leftarrow i_{\text{em}} + 1$
28:　　　**if** $i_{\text{em}} \geq N_{\text{EM max}}$ **then**
29:　　　　　**break**
30:　　　**end if**
31:　　**end while**
32:　　**return seg**, $\mathbf{M}$, $\mu$, $\sigma$
33: **end function**

---

transform on the inverted binary image, where the arterial structure (originally foreground, intensity 1) is inverted to background (0), and non-vessel regions (originally 0) become foreground (1). This allows bwdist to calculate the straight-line (Euclidean) distance from each voxel inside the vessel to the nearest vessel wall voxel. The output is a distance map where each voxel's value represents its Euclidean distance to the nearest background voxel with an index map identifying the linear index of the corresponding nearest background voxel.

3. **Skeletonization:** The arterial structure is reduced to a one-voxel-thick centerline through 3D skeletonization, preserving the topological connectivity and branching structure of the arterial network. This centerline represents the medial axis of the vessels, serving as the reference for radius measurements. For the purpose of this study, iterative thinning technique was employed for skeletonizing the 3D tubular structures that ensured the centerline accurately reflects the vessel's geometry Kollmannsberger et al. (2017); Lee et al. (1994).

4. **Radius Computation:** The distance transform map and skeletonized 3D structure is combined to compute radius for centerline points. For each voxel on the centerline, the algorithm retrieves its 3D coordinates and identifies the nearest background voxel from the index map generated by the distance transform. The distance between the centerline voxel and its nearest background voxel is calculated, adjusted for the anisotropic voxel dimensions. For a centerline voxel at position $(x_i, y_i, z_i)$ with the nearest background voxel at $(d_x, d_y, d_z)$, the radius is computed as:

$$r = \sqrt{\begin{array}{l}((x_i - d_x) \cdot \text{unit.x})^2 + \\ ((y_i - d_y) \cdot \text{unit.y})^2 + \\ ((z_i - d_z) \cdot \text{unit.z})^2\end{array}} \qquad (9)$$

This radius value represents the shortest path from the centerline close to the vessel wall, providing a precise measure of the vessel's radial extent at that point. Supplementary Fig. S1 shows the distance transform (in pixels) maps computed for 3 different test cases in 2D.

5. **Centerline Output:** The computed radii represented and stored in a 3D variable, with non-zero values assigned only to centerline voxels takes extensive memory and processing time. Sparse representation of this data is used for all further processing and vessel fusing algorithm.

*Implementation Details*

The algorithm is designed for computational efficiency and accuracy, leveraging vectorized operations and optimized libraries for distance transform computation and skeletonization. The distance transform, inspired by the functionality of bwdist, is particularly effective for 3D medical imaging datasets, as it scales well with image size and handles complex geometries. The incorporation of unit corrections ensures that radial measurements are physically accurate, addressing challenges posed by non-uniform voxel spacing in modalities such as Magnetic Resonance Angiography (MRA). The resulting centerline and diameter data are stored in a sparse format compatible with downstream graph generation processes.

### 2.1.3. Create & Label Vessel-Fused Network Nodes
*Vessel-Fused Network Generation*

In general, when generating a graph from a 2D/3D data, the primary objective is to identify the objects and then the relationships between them. In the case of TOF MRA arteries, nodes/landmarks are the objects, and the traces are the relationships. In this work, we have implemented a novel approach that works with 3D morphologicaly volumes representing them as sparse data and yields in unique one-to-one vessel-fused graph with respective traces, nodes and hubs (Fig. 2, Algorithm 3 outlines this iterative process). The located hubs finally decompose into a single node that is equidistant from all its endpoint nodes. The vessel-fused graph construction ensures complete connectivity. To construct a reliable vessel-fused graph, we need to ensure robustness in node and trace identification, thus a 3d viewer landmarking tool is developed and publicly hosted for that purpose. Merging the user input this constructs a comprehensive graph of the arterial network, dividing the chunks with proper labels.

The key components for the Algorithm 3 is as follows-

1. **Nodes, Hubs, and Traces:** Voxels are categorized by connectivity: degree-1 voxels are endpoints, degree-2 voxels are between-points, and degree-3 voxels are nodes or part of a hub. A hub is a cluster of degree-3 voxels with a central node equidistant from all associated endpoints. 3D traces are continuous arterial segments formed by connected between-points.

   An example MIP skeleton illustrating the spatial distribution of initial nodes (red), hubs (blue), and their corresponding hub-endpoints is shown in Fig. 3.

2. **Trace Determination:** A spatial map identifies Nodes and Hubs. For each Hub, the central node, approximately equidistant from endpoints, is calculated. All possible Traces connecting Nodes are then computed to define arterial pathways using Algorithm 3.

3. **Dynamic Graph Table Creation:** A dynamic graph table is generated, incorporating user inputs from the node labelling guided with hessian map. This table adapts to assigned node labels, providing a customized arterial graph for further analysis.

*Node Labelling with Hessian Guide*

The Labelling Guide helps the user to identify potential nodes sequentially. Arteries that are very close to each other may sometimes be connected, so edges can be deleted on this panel to remove spurious connections. When the user saves completed labels, the next step is to upload the labels to the toolbox, which computes the final arteries and artery features.

---

**Algorithm 3** Vessel-Fuse Network Creation

1: **function** VESSELFUSEDNETWORKCREATION(Skeleton3D)
2:　　Endpoints, Nodes, Hubs ← IdentifySkeletonComponents(Skeleton3D)　　▷ Identify endpoints, nodes, hubs
3:　　Initialize VesselFusedNetwork with empty GlobalNodes, GlobalTraces
4:　　**for** each Hub in Hubs **do**
5:　　　　MiniSubgraph ← CreateMiniSubgraph(Hub)　　▷ Create subgraph for hub
6:　　　　HubCenterNode ← IdentifyHubCenterNode(MiniSubgraph)　　▷ Identify central node
7:　　　　HubSubTraces ← IdentifyHubSubTraces(MiniSubgraph, HubCenterNode)　　▷ Paths from endpoints to center
8:　　　　VesselFusedNetwork.GlobalNodes ← VesselFusedNetwork.GlobalNodes ∪ MiniSubgraph.Nodes
9:　　　　VesselFusedNetwork.GlobalTraces ← VesselFusedNetwork.GlobalTraces ∪ HubSubTraces
10:　　**end for**
11:　　**return** VesselFusedNetwork
12: **end function**

---

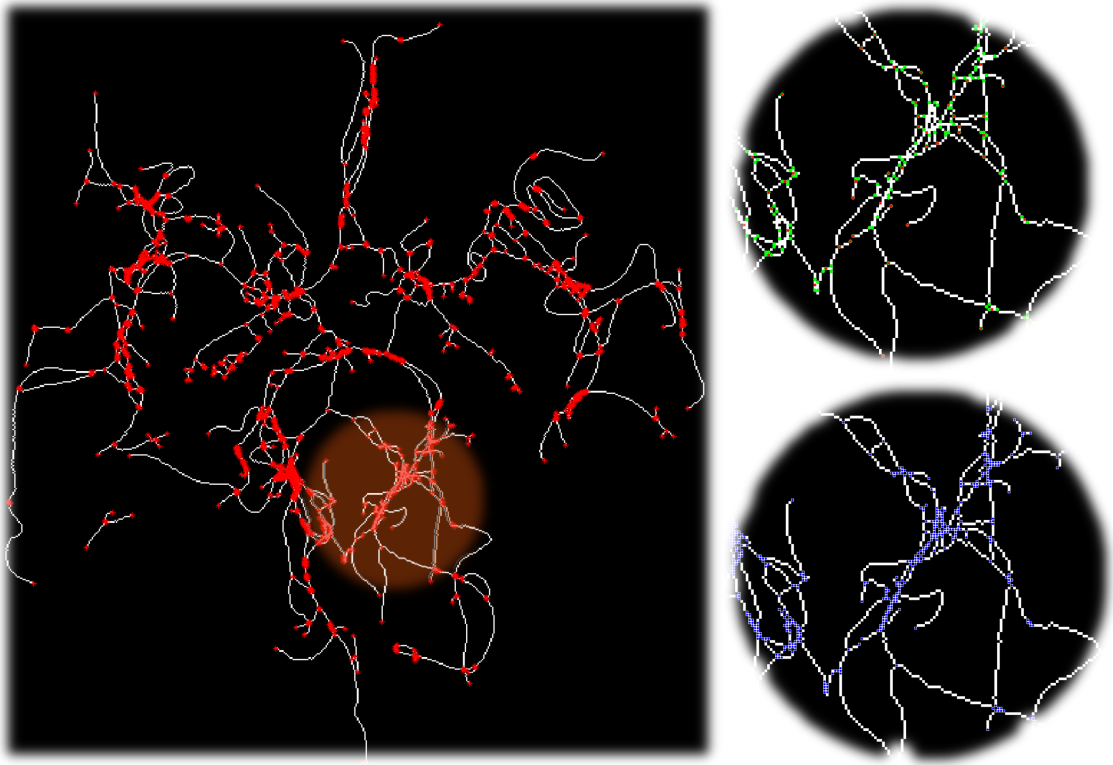

Fig. 3: Example maximum intensity projection (MIP) skeleton demonstrating spatial distribution of initial nodes (red), and hubs (blue), and the corresponding hub-endpoints (green).

### 2.1.4. Artery Feature Extraction

For each classified artery and its graph, we calculate the following features:

a. Total length is computed by summing lengths of all valid intracranial artery segments.

b. Radius already sampled in step 3 before rasterization.

c. Total volume refers to the point-by-point volume computation using truncated cone formula.

d. Total number of branches refers to the count of arterial segments between bifurcation points or between a bifurcation and a terminal branch.

e. Mean section area is computed along the tracing points, reported as the mean of overall trace.

f. Mean Surface area is the area over given trace points.

g. Tortuosity is computed as the ratio of an artery length and the Euclidean distance between its start and end points.

h. Fractal dimensionality is calculated using the box-counting method (Madan and Kensinger, 2016), providing a quantitative measure of the morphological complexity in cerebral arteries.

### 2.2. Simulation Strategy of ArteryX

One of the goals of our simulation framework is to generate arteries that preserve the characteristics and organization found in the human brain when simulating ground truth values for different features (Fig. 4). To achieve this goal, our approach leverages the landmark graph derived from an in-vivo subject. ArteryX's evaluation framework is used to make primary landmark extraction possible. To mitigate toolbox-related bias in our ground-truth data simulation, we have represented each artery as a 1D signal and defined the required ground truth values independently during rasterization. The only information that is used from the ArteryX's evaluation framework is the landmark (i.e., nodes) of starting and end point of each artery. Each 1D artery signal is then oriented with a random rotational factor and rasterized into a 3D volume using the ground truth radius distribution across different tracing points. Ground truth values for tortuosity and artery length are computed before the radius-based rasterization is performed. If the registration parameters to MNI data are available, the landmarks can be independently defined using the MNI spatial probability distribution of the arteries.

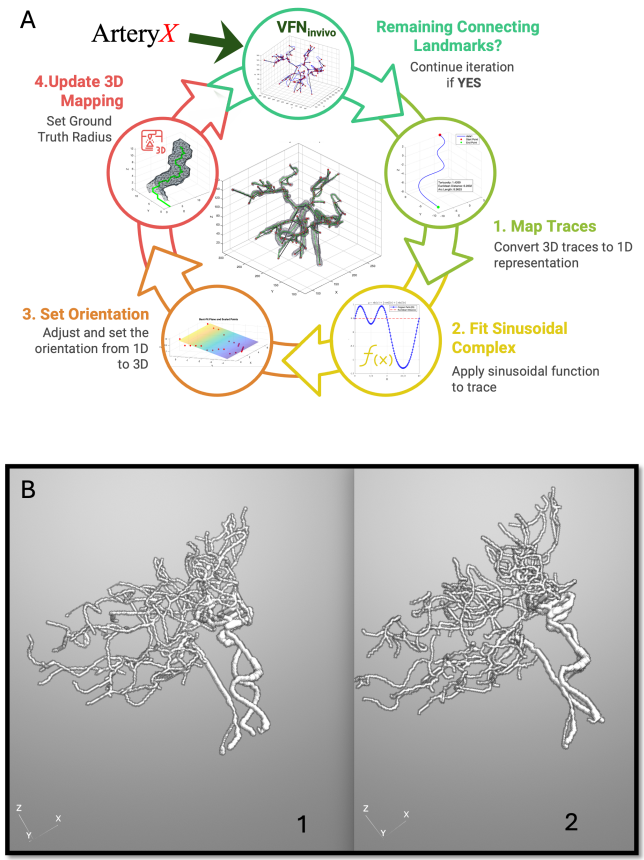

**Fig. 4:** Simulation Strategy using ArteryX derived Vessel Fused Network (VFN) from an *in-vivo* MRA: A. The Vessel Fused Graph provides the end-point landmarks for each artery type along with its radius profile, tracing, and orientation. Synthetic Ground Truth radius profile, orientation and tracing is then applied on the connecting end points iteratively to obtain the final simulated data. B. Using the same VFN but different orientations, two different artery profiles are generated and shown as examples.

A Fourier basis dictionary is created to serve as a platform for the simulation framework for storing and sharing different types of arteries that need to be analyzed during simulation. In short, to simulate a specific type of artery, the following steps are taken:

1. **Landmark Extraction from in-vivo:** The landmarks of start and end points of the arteries can be extracted using the ArteryX evaluation framework (Algorithm 1) with the vessel-fused graph. The evaluation framework computes 3D tracing and uses a novel algorithm (Algorithm-3) to figure out the core nodes, enabling the vessel-fused network creation. ArteryX maintains a dictionary of the landmarks of the vessel-fused graphs with the key tracings.

2. **Representing in-vivo Arteries as a 1D Signal**
   ArteryX makes use of representing arteries as 1D signals to instantiate the ground truth parameters in creating them, while the landmarks create a coherent global artery topology. These 1D signals are represented using a Fourier basis dictionary (FBD) to make the sampling number independent when reconstructing the signal. The Fourier basis makes the reconstruction and sharing of arteries easier. For example, the FBD enables the creation of synthetic artery data using different 1D arteries from different in-vivo subjects. The process of creating traces from in-vivo subjects is elaborated in Section 2.1.3. It highlights the use of a 2D orientation plane for each of the arteries. As a result, the simulation framework can generate new synthetic datasets based on the modifications of orientation, landmark, and radius profile.

3. **Artery Computation**

   a. **1D Fourier Basis to 3D Traces with Varying Orientation:** Artery computation begins with a landmark graph derived from in-vivo data. If the data is not registered to MNI space, the artery probability space constraint is not applied, and constant landmarks are used.

   b. **3D Centerline to Rasterization with Radius Distribution:** Artery signals are synthesized from the FBD by random sampling and orienting, and finally rasterized using an in-house rasterization algorithm with radius distribution for a known artery group, thus creating a unique dataset.

4. **Ground Truth Features** For each connecting landmark points, we use two user-defined features to simulate our artery segments.

   a. Orientation of the trace points with a 3d fitted plane.

   b. Artery radii for each trace points.

   Following features are then computed based on the predefined orientation, radius and centerline-points.

   a. Point-by-point volume computation with a truncated cone formula.

   b. Point-by-point Section Area and its mean.

   c. Surface area.

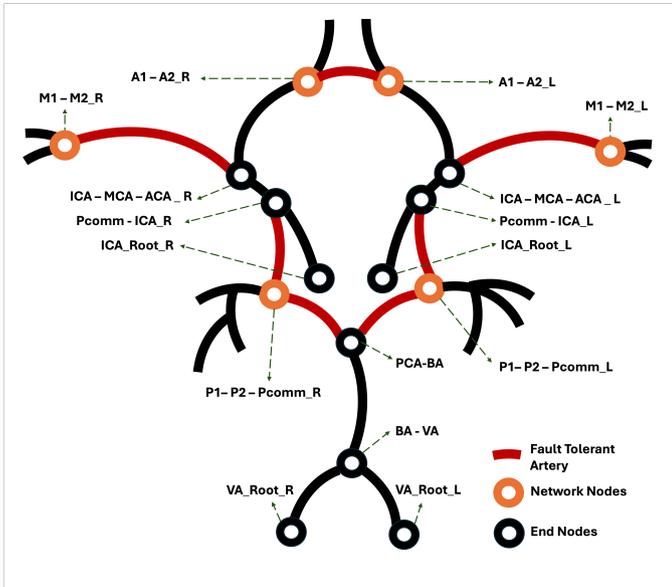

Fig. 5: ArteryX evaluation pipeline illustrating the arterial network with 16 critical nodes. Red edges denote *fault-tolerant arteries* that may be absent, while orange circles represent network nodes, and black circles indicate end nodes. Dashed lines highlight key connections between nodes.

d. Tortuosity is computed from the length and using start and end points of the artery.

e. Fractal dimensionality

### 2.3. Standardized and Generalizable Artery Classification

To establish a standardized and adaptable framework for the intracranial arterial system, arteries are classified into two main networks: Proximal Arteries and Distal Arteries. The latter are subdivided into subnetworks to support depth-based measurements and feature extraction as needed. Supplementary Table S1 provides a comprehensive list of arteries and their subnetworks with detailed descriptions. This classification underpins the dynamic graph table (see Section 3), which accommodates variations, including the potential absence of *fault-tolerant arteries* marked in red (Fig. 5).

The ArteryX toolbox 3D viewer requires the annotation of 16 critical nodes for accurate classification and adaptability across diverse arterial profiles. These nodes, selectable within the viewer, are:

- **M1-M2_L, M1-M2_R**: Middle cerebral artery segment connecting M1 to M2 (applicable to both left and right sides).

- **A1-A2_L, A1-A2_R**: Anterior cerebral artery segment connecting A1 to A2 (applicable to both left and right sides).

- **ICA-MCA-ACA_L, ICA-MCA-ACA_R**: Internal carotid artery bifurcating to middle and anterior cerebral arteries (applicable to both left and right sides).

- **Pcomm-ICA_L, Pcomm-ICA_R**: Posterior communicating artery connection to internal carotid artery (applicable to both left and right sides).

- **ICA_Root_L, ICA_Root_R**: Internal carotid artery root (applicable to both left and right sides).

- **P1-P2-Pcomm_L, P1-P2-Pcomm_R**: Posterior cerebral artery segment connecting P1, P2, and posterior communicating artery (applicable to both left and right sides).

- **PCA-BA**: Posterior cerebral artery connection to basilar artery.

- **BA-VA**: Basilar artery connection to vertebral artery.

- **VA_Root_L, VA_Root_R**: Vertebral artery root (applicable to both left and right sides).

The dynamic graph table is formulated to account for the potential absence of *fault-tolerant arteries*, ensuring robustness in arterial network analysis. The network node ensures to report distal arteries (such as MCA, ACA and PCA) summarizing them as graphs.

### 2.4. Healthy vs. CSVD Group Comparison

Following institutional guidelines and with written informed consent, participants were recruited from two distinct studies: one conducted in 10 healthy individuals examining the impact of varying imaging resolutions and another with 20 participants to assess differences between those with and without cerebral small vessel disease (CSVD). Participants were selected based on the presence or absence of white matter hyperintensity (WMH) lesions, a hallmark of CSVD. The study included ten healthy controls without CSVD (CSVD-) (age range: 51–69 years, mean ± SD: 56.4 ± 8.2; 3 females; mean lesion volume ± SD: 0.05 ± 0.11 cm$^3$) and ten individuals with CSVD (CSVD+) (age range: 51–73 years, mean ± SD: 64.2 ± 6.5; 4 females; mean lesion volume ± SD: 1.3 ± 2.5 cm$^3$).

#### 2.4.1. Image Acquisition

MR Imaging was performed using a Siemens 3T MAGNETOM Prisma Fit scanner (Erlangen, Germany), equipped with a 64-channel head coil (receive-only), body coil for transmission, and high-performance gradients (maximum strength: 80 mT/m, slew rate: 200 mT/m/s). The MRI protocol comprised T1-weighted 3D MPRAGE imaging (TI: 962 ms, TE/TR: 2.34 ms/1840 ms, 1 mm isotropic resolution) for anatomical reference, along with TOF MRA sequences (TE/TR: 3.42 ms/21 ms, flip angle: 18°, resolutions: 0.52 × 0.52 × 0.8 mm$^3$ and 0.26 × 0.26 × 0.5 mm$^3$). The images were checked for any artifacts including motion, or signal dropout.

#### 2.4.2. Statistical Analysis

Statistical analyses were conducted using MATLAB (version 2023b). Vascular features were compared using bar plots and percent difference plots for several arteries, comparing ArteryX and iCafe against synthetic ground-truth values. Pearson correlation coefficients were computed to assess the association between the two methods (ArteryX and iCafe), as well as their respective correlations with the ground-truth. For cohort comparisons, vascular features extracted from healthy and CSVD subjects were compared using unpaired t-tests for several arteries, separately for ArteryX and iCafe. A $p-value < 0.05$ was considered statistically significant.

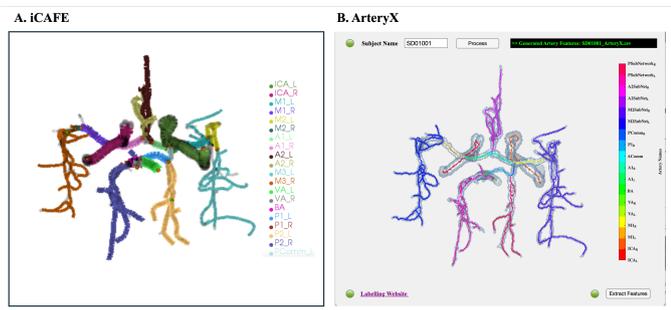

Fig. 6: Example artery classifications from iCafe and ArteryX

## 3. Results

The performance of the proposed ArteryX framework is directly compared with iCafe, another semi-supervised approach, as expert-guided semi-supervised approaches — particularly those informed by artery anatomy specialists or radiologists - can substantially enhance the accuracy of artery landmarking and classification. Both ArteryX and iCafe were independently subject to ground truth validations for important topological, morphological and complexity features. The independent validation involved curating synthetic data from the ArteryX's synthetic data generation framework which utilizes Vessel Fused Graph (Fig. 2) obtained from an in-vivo data. The data generation framework shows (Fig. 4) how different arterial profiles can be generated utilizing the graph network from an in-vivo data and defining artery features (e.g., orientation, radius) from scratch.

Fig. 5 depicts the arterial network consisting of 16 nodes from the ArteryX evaluation framework. Fig. 6 presents the discrete artery classification results from both ArteryX and iCafe on a synthetic data. As illustrated in Fig. 7, comparisons with ground-truth synthetic data indicate that ArteryX produces artery feature estimates that closely aligned with those of iCafe across most local arterial regions. ArteryX showed only a 0–10% deviation from the ground-truth in key morphological metrics, including total length, tortuosity, mean radius, and surface area — significantly outperforming iCafe, which exhibited larger discrepancies. Although ArteryX tended to overestimate mean section area in most arteries and slightly overestimated volumes in the proximal, distal, MCA, and ACA regions, its overall accuracy and consistency remained superior to that of iCafe (Fig. 7B). Similar result is observed for topological branch counts, where iCafe shows higher deviation by underestimating the number of branches particularly in the distal arteries. Additionally, ArteryX reports complexity features like fractal dimensionality of the discrete arteries with less than 10% error; which is not available in any publicly available toolboxes to the best of our knowledge.

Fig. 8 compares the outcomes of iCafe and ArteryX in terms of Pearson correlation for both synthetic ground-truth and in-vivo data. ArteryX demonstrates highly significant and very strong correlations with the synthetic ground truth across all vascular features ($r > 0.9, p < 0.0001$). In contrast, iCafe shows only low to moderate correlations for tortuosity ($r = 0.66, p < 0.05$) and surface area ($r = 0.57, p > 0.05$), although significant correlations were observed for other features.

The observation from the synthetic data aligns with the correlation of ArteryX results with iCafe. The trend is observed in the in-vivo data (Fig. 8B1,B2), confirms that ArteryX consistently outperforms iCafe across all the vascular features, particularly in the case of Surface Area.

Further, group comparisons between healthy and CSVD cohorts indicate that ArteryX exhibits greater sensitivity in detecting vascular differences across several arteries—particularly in surface area, total length, and total branch counts. In contrast, iCafe shows a significant difference only in the total length of the PCA (Table 1). The differences in the distal artery is not captured with iCafe but ArteryX. Supplementary Fig. S2 shows the relationship between age and the fractal dimensionality of the middle cerebral artery in individuals with and without CSVD. A significant positive correlation is observed in the subjects with CSVD group ($r = 0.65, p = 0.042$), while the healthy group shows a non-significant trend ($r = 0.47, p = 0.166$), suggesting a stronger age-related increase in vascular complexity in CSVD.

ArteryX solves the dangling and disconnected artery problem in the iCafe which requires substantial time for the operator to finalize the landmarking. which enabled reliable extraction of the arteries and reducing operator variability. ArteryX was tested on Mac and Linux systems which demonstrated significantly faster performance. The only user input required was landmarking, which took approximately 8 to 10 minutes per subject. In contrast, iCafe required an average of 1.5 hours per subject for both landmarking and the correction of dangling or disconnected arteries.

To enhance visualization and understanding of the arteries, Supplementary Fig. S3 demonstrates land-marking guides with a virtual reality (VR)-enabled 3D atlas. The relevant arteries are described in Supplementary Table S1.

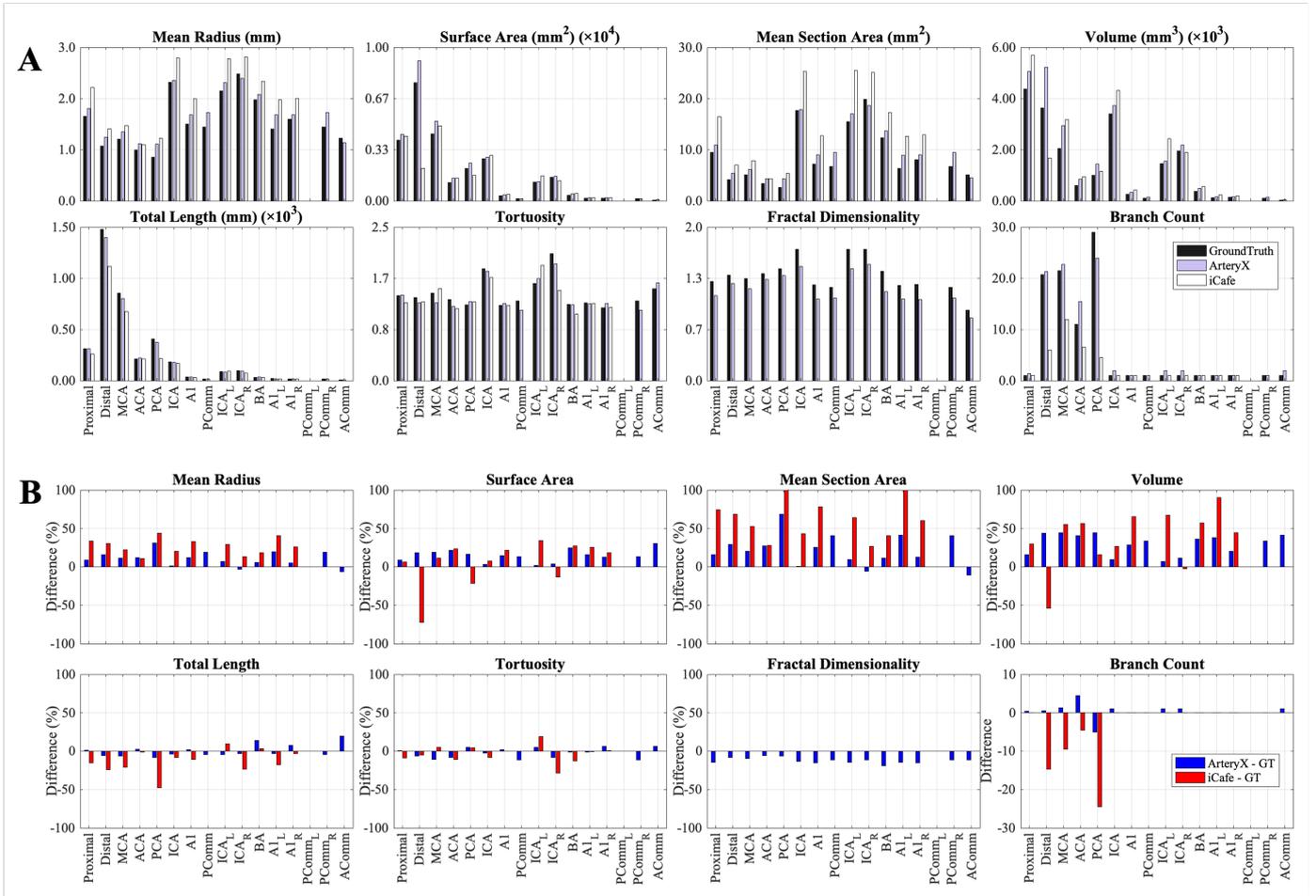

Fig. 7: Validation of vascular features across multiple arteries for ArteryX using ground-truth synthetic data and comparison with semi-automated iCafe. A. Bar plots comparing vascular features from ArteryX with ground-truth synthetic data and iCafe, showing ArteryX outperforms iCafe; B. Percent differences between ground-truth and both ArteryX and iCafe. Full artery names are listed in Supplementary Table S1.

## 4. Discussion

In this study, we present a comprehensive framework for validating and extracting 3D brain artery features using a graph-based approach. This work addresses several longstanding gaps in the field — firstly, the lack of a robust standardized classification system of arteries and validation framework that captures the anatomical complexity and variability found in the in-vivo artery profiles by uniquely leveraging vessel-fused networks.

Secondly, the semi-supervised approaches have shown good accuracy in artery classification but struggle with tracing issues, particularly dangling or disconnected arteries (Chen et al., 2020). Resolving this typically requires manual correction and landmarking by a trained expert, taking around 1.5 hours for 0.52x0.52x0.5 mm³ resolution TOF-MRA data. In the ArteryX evaluation framework, we have automated the process and reduced the processing time 6 folds.

Thirdly, previous studies on semi-automated or automated toolboxes have primarily focused on segmentation dice scores, often missing the important evaluation and validation of specific vascular features. In cases where features like radius or length were assessed, they relied heavily on hand-labeled ground-truth from radiologists. However, such annotations are subject to variability(Zampakis et al., 2015; Mirza et al., 2024) due to anatomical variations, imaging modalities, and non-standardized measurement techniques. Also, 1D annotations are often limited in scale and anatomical diversity. In contrast, we propose to shift this evaluation criteria focusing toward the accuracy of artery-specific feature extraction, using synthetic data with known ground-truth for robust validation.

Finally, automated deep learning techniques for artery classification often achieves less than 90% accuracy, despite requiring training on hundreds of subjects. This limits their practicality in clinical and research settings. Moreover, models are typically tailored to specific scanners or imaging parameters, making retraining for new datasets time-consuming and resource-intensive. Our simulation strategy addresses these challenges by harnessing vessel-fused network derived from a small number of in-vivo subjects to generate hundreds of different artery profiles. By simulating specific landmark annotations from scratch as validation sources and synthetic arterial data with strong anatomical resemblance to human cerebral vasculature, our framework provides a valuable benchmark and protocol-specific arterial simulation for required TOF-MRA field-of-view.

The dynamic generation of ground-truth data — reflecting re-alistic topology, morphology, and branching patterns - enables high-fidelity validation across diverse clinical populations and imaging conditions. Our graph-based approach preserves arterial topology, allowing segmented vessels to be meaningfully divided into hierarchical chunks while maintaining continuity and vessel identity. This topological preservation is critical for downstream morphometric analyses, especially when studying disease-associated patterns in distal branches.

Table 1: Comparison of vascular features in human subjects with and without CSVD using ArteryX and iCafe. CSVD+ indicates presence of CSVD; CSVD- indicates absence. Significant arterial vascular features are shown. Units: mm (length), mm² (area), mm³ (volume) where applicable.

| Significant Variables | CSVD+ Mean | CSVD- Mean | Comparison | p_value | t_value |
|---|---|---|---|---|---|
| **Table-1A:** *ArteryX* | | | | | |
| Distal- Surface Area | 8046.1 | 10590 | CSVD+ < CSVD- | **0.043947** | -2.2317 |
| Distal- Total Length | 1616.3 | 2149.4 | CSVD+ < CSVD- | **0.035366** | -2.3277 |
| Distal- Total Branch Count | 23.029 | 31.933 | CSVD+ < CSVD- | **0.0077402** | -3.0007 |
| MCA- Surface Area | 5906.3 | 7248.1 | CSVD+ < CSVD- | **0.045793** | -2.1471 |
| MCA- Total Length | 1200.8 | 1479.5 | CSVD+ < CSVD- | **0.043891** | -2.1671 |
| MCA- Total Branch Count | 26.45 | 33.45 | CSVD+ < CSVD- | **0.025523** | -2.44 |
| **Table-1B:** *iCafe* | | | | | |
| PCA_L- Length | 118.68 | 177.64 | CSVD+ < CSVD- | **0.0087385** | -2.9548 |

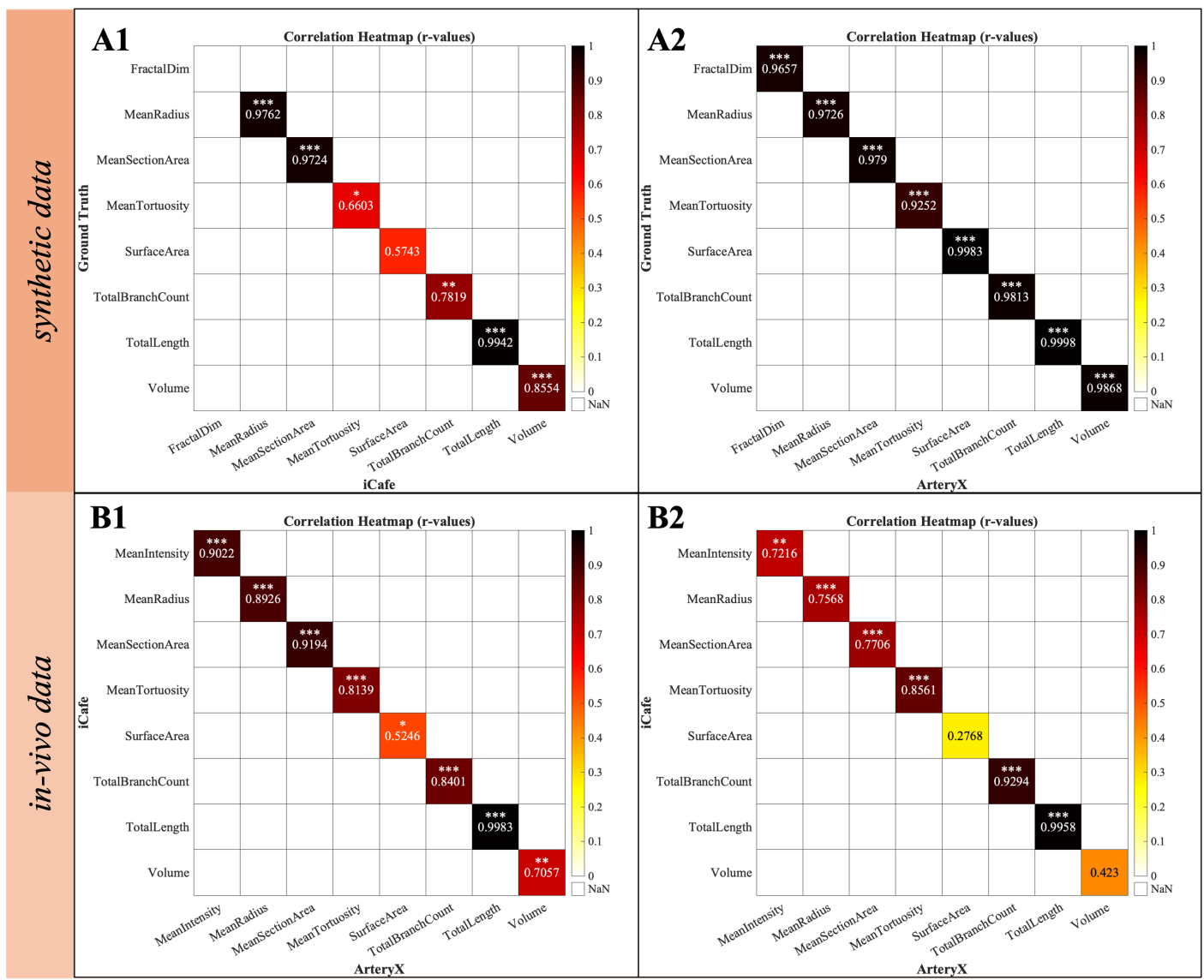

Fig. 8: Correlation of ArteryX vs iCafe for synthetic data and *in-vivo* is shown. A1 and A2 are showing the correlation with Ground Truth for iCafe and ArteryX respectively. B1 and B2 respectively results from high resolution and lower resolution *in-vivo* TOF MRA data. ($p < 0.05$, $p < 0.001$ and $p < 0.0001$ respectively indicated with *, ** and ***)

Another novel component of our work is the use of fractal dimensionality and morphological features (e.g., volume, surface area) to quantify distal artery characteristics. These metrics provide richer descriptors of vascular complexity and are particularly relevant for analyzing small vessel alterations in conditions like CSVD.

One of the primary advantages of our framework is its efficiency and practicality. Each step of the pipeline is built with sparse computational operations, enabling significant memory savings without compromising accuracy. The result is a streamlined, semi-automated approach to artery quantification that requires minimal user intervention—up to six times less processing time than current tools like iCafe. Furthermore, our method offers a rich set of morphological features from segmented arteries, providing a broader context for neurovascular studies compared to existing solutions.

Despite these strengths, ArteryX framework has limitations. While it identifies missing arteries like anterior and posterior communicating arteries, more complex anatomical variations (e.g. stroke) will be incorporated in future versions. Looking ahead, ArteryX's robust ground-truth modeling can support large synthetic data generation, by incorporating GAN-based simulations of in-vivo MRA data, and providing high-quality labeled data - crucial for training deep learning models that often struggle with data scarcity and annotation variability.

## 5. Conclusions

We have standardized the classification of the arteries from TOF-MRA, and presented a framework for validation and semi-automated feature retrieval for 3D brain artery analysis using vessel-fused graphs. This is the first framework to reduce user-interaction by leveraging graph nodes for artery tracing and labeling. The study utilizes synthetic MRA simulated from the framework and user-defined ground-truth features to validate feature estimation accuracy, establishing a benchmark for toolbox comparison and has the potential to reducing the need for acquiring large training datasets in deep learning applications. Furthermore, *ArteryX* derived features demonstrated increased sensitivity compared to state-of-the-art iCafe toolbox in both simulation and in-vivo comparisons across healthy and CSVD cohorts. The toolbox requires minimal user input and is six times faster in processing intracranial arteries, demonstrating great potential for clinical and population-based studies.

## Declaration of competing interest

The authors declare that they have no competing financial interests or personal relationships that could have influenced the work presented in this article.

## Acknowledgments

This study was supported by a grant from the National Institutes of Health (R03NS134395-01 and R01MH118020). The authors sincerely thank the study participants and coordinators for their invaluable contributions.

## Appendix A. Supplementary Material

The following is the supplementary material related to this manuscript.

**Table S1.** *Intracranial artery segments abbreviation table.*
**Figure S1.** *Distance Transform maps in pixels is shown for three different 2D tube shaped structures.*
**Figure S2.** *Association between age and MCA fractal dimensionality in CSVD+ and CSVD- groups.*
**Figure S3.** *Interactive VR environment for cerebral artery learning and labeling.*

## Availability of research data/code

The toolbox is publicly available on GitHub at abrar-faiyaz.github.io/arteryX/. The code and synthetic data will be made available upon request.